\documentclass[12pt,preprint]{aastex}

\slugcomment{Accepted by  ApJ Letters on October 6, 2014}

\shorttitle{Holm~15A: the  Largest Core  Known so far}
\shortauthors{L\'opez-Cruz et al.}

\begin{document}


\title{The Brightest Cluster  Galaxy in Abell~85: The Largest Core  Known so far}


\author{O. L\'opez-Cruz\altaffilmark{1,7}, C.  A\~ norve\altaffilmark{1,2},  M. Birkinshaw\altaffilmark{3}, D.~M.,  Worrall\altaffilmark{3}, H.~J.  Ibarra-Medel\altaffilmark{1},  W.~A. Barkhouse\altaffilmark{4,7},  J.~P. Torres-Papaqui\altaffilmark{5},   V. Motta\altaffilmark{6} }





\altaffiltext{1}{Instituto Nacional de Astrof{\'\i}sica, \'Optica y Electr\'onica (INAOE),  Astrof{\'\i}sica,
Luis Enrique Erro No.1,  Tonantzintla, Pue., C.P. 72840, M\'exico. e-mail: omarlx@inaoep.mx}
\altaffiltext{2}{FACITE, Universidad Aut\'onoma de Sinaloa,
Blvd. de la Americas y Av. Universitarios S/N, Ciudad Universitaria, C.P. 80010, Culiac\'an
Sinaloa, M\'exico}
\altaffiltext{3}{HH Wills Physics Laboratory, University of Bristol, Tyndall Avenue, Bristol, BS8 1TL, UK}
\altaffiltext{4}{Department of Physics \& Astrophysics, University of North
Dakota, Grand Forks, ND 58202, U.S.A.}
\altaffiltext{5}{Departamento de Astronom{\'\i}a, Universidad de Guanajuato (DAUG), Callej{\'o}n Jalisco S/N Col. Valenciana, C.P. 36240, Guanajuato, Gto., M\'exico. }
\altaffiltext{6}{Instituto de F{\'\i}sica y Astronom{\'\i}a, Universidad de Valpara{\'\i}so, Avda. Gran Breta\~na 1111, Vara{\'\i}so, Chile.}
\altaffiltext{7}{Visiting Astronomer, KPNO is operated by AURA, Inc.\ under contract to the NSF.}


\begin{abstract}
We have found that the brightest cluster galaxy (BCG) in Abell~85,
\objectname[A85-BCG]{Holm~15A}, displays the largest core so far
known. Its cusp radius, $r_{\gamma} = 4.57 \pm 0.06$ kpc ($4\farcs26
\pm 0\farcs06$), is more than 18 times larger than the mean for
BCGs, and $\gtrsim 1$~kpc larger than \objectname{A2261-BCG}, 
hitherto the largest-cored BCG \citep{PL12}. 
Holm~15A hosts the luminous amorphous radio source 0039-095B and
has the optical signature of a LINER. Scaling laws
indicate that this core could host a supermassive black hole (SMBH) of mass
$\mathrm{M}_{\bullet}\thicksim (10^{9}-10^{11})\,\mathrm{M}_{\sun}$. We suggest
that cores this large represent a relatively short phase in the evolution of BCGs, whereas
the masses of their associated SBMH  might be set by initial conditions. 
\end{abstract}






\section{Introduction}

We have been aware of the enormous core of Holm~15A \citep{Ho37}, the
brightest cluster galaxy (BCG) in Abell~85, for over thirty years.
\citet{Ho80} fitted a modified Hubble law to its surface-brightness
profile, and reported a core radius (i.e., the radius where the
surface brightness reaches half of its central value)  
$r_{c} = 5\farcs72$, which corresponds to a
physical size $r_{c}=6.14$ kpc at Abell~85's restframe
\citep[$z_{clus} = 0.05529 \pm 0.00024$; combined redshift,
data taken from the literature and our analysis of data from SDSS
DR9;][]{DR9}. Extended ${\rm H}_{\alpha}$+[\ion{N}{2}] and
[\ion{O}{3}] $\lambda 5007$ emission was reported by
\citet{HCW85} and \citet{FIF95}. \citet{JPB97} using LOCOS data \citep{L00} found that Holm~15A has a large core.
Although \citet{Fa97} had already suggested the presence of
cores in galaxies brighter than $M_{V} =-21$ mag, establishing
the core-cusp dichotomy, this  was largely
overlooked due, in part, to the lack of resolution
that had hampered most BCG studies, and the additional
complication of using  different  parameterizations.  As a
result, Holm~15A has remained a curiosity.    
 

Before the distribution of cores was established \citep[e.g.,][]{La03,La07a}, \citet{Fa97} had advanced the idea that 
large cores were a manifestation of  SMBH  binaries.  Although, we are still  unclear about details such as BH merging times, hardening, and the so-called ``final parsec problem'' \citep[e.g.,][]{Ka13}, the view that cores are expanded  by the ``scouring'' action of SMBH over galactic cusps has  gained wide
acceptance \citep[e.g.,][]{Me06, KH13}.  Nevertheless, there are alternative scenarios, for example,  scalar-field dark matter  \citep[e.g.,][]{RM12}, in which cores are formed without  secondary mechanisms.  

\citet{A12} and A\~norve \& L\'opez-Cruz (2014, in preparation)  have
conducted a comprehensive study of the structure of galaxies using  
LOCOS clusters at $z\leq 0.08$, using the Driver for GALFIT
on Cluster Galaxies \citep[DGCG,][]{A12}, a Perl script for GALFIT \citep{Peng10} that accounts for the effects of
crowding and point spread function (PSF) variations. DGCG
allows the 2-D modeling (bulge+disk) of the surface-brightness distribution for cluster galaxies.   
Holm~15A was one of the few galaxies for which our DGCG analysis did 
not converge. This prompted us to perform the more detailed study 
that we report in this Letter.
 
We use the  work of \citet[][hereafter PL12]{PL12} and
\citet[][and references therein]{La07a} to place Holm~15A into the
context of the overall BCG population. To allow direct comparisons with PL12, 
we adopted $\Omega_{\mathrm{m}}=0.3$,
$\Omega_{\Lambda}=0.7$, and $\mathrm{H}_{0}=70\;{ \rm km\,s^{-1}\,Mpc^{-1}}$.

\section{Observations}

\subsection{Optical Observations}
We work from a  LOCOS image of Holm~15A.
This image was taken under good weather conditions with an average
$1\farcs67$ seeing (pixel scale $ = 0\farcs68/\mathrm{pixel}$).  To
supplement our results, we  looked for higher-resolution images. We
failed to find HST images. However, Abell~85 is part of the ongoing
Multi-Epoch Nearby Cluster Survey \citep[MENeaCS,][]{Sa11}, from which
we selected an image with 120-s integration in the SDSS
$r^{\prime}$ band, taken with the CFHT 3.5m-telescope and 
MegaCam (see Table~1). 

The LOCOS image was reduced  with IRAF,  according to standard
reduction procedures \citep[e.g.,][]{L04}.  The MENeaCS were 
processed using the Elixir pipeline \citep{Sa11}.  Optical spectra
from SDSS DR9  were used to determine cluster membership, galaxy and
cluster velocity dispersions, and line ratios.

\begin{figure}
\epsscale{0.95}
\plotone{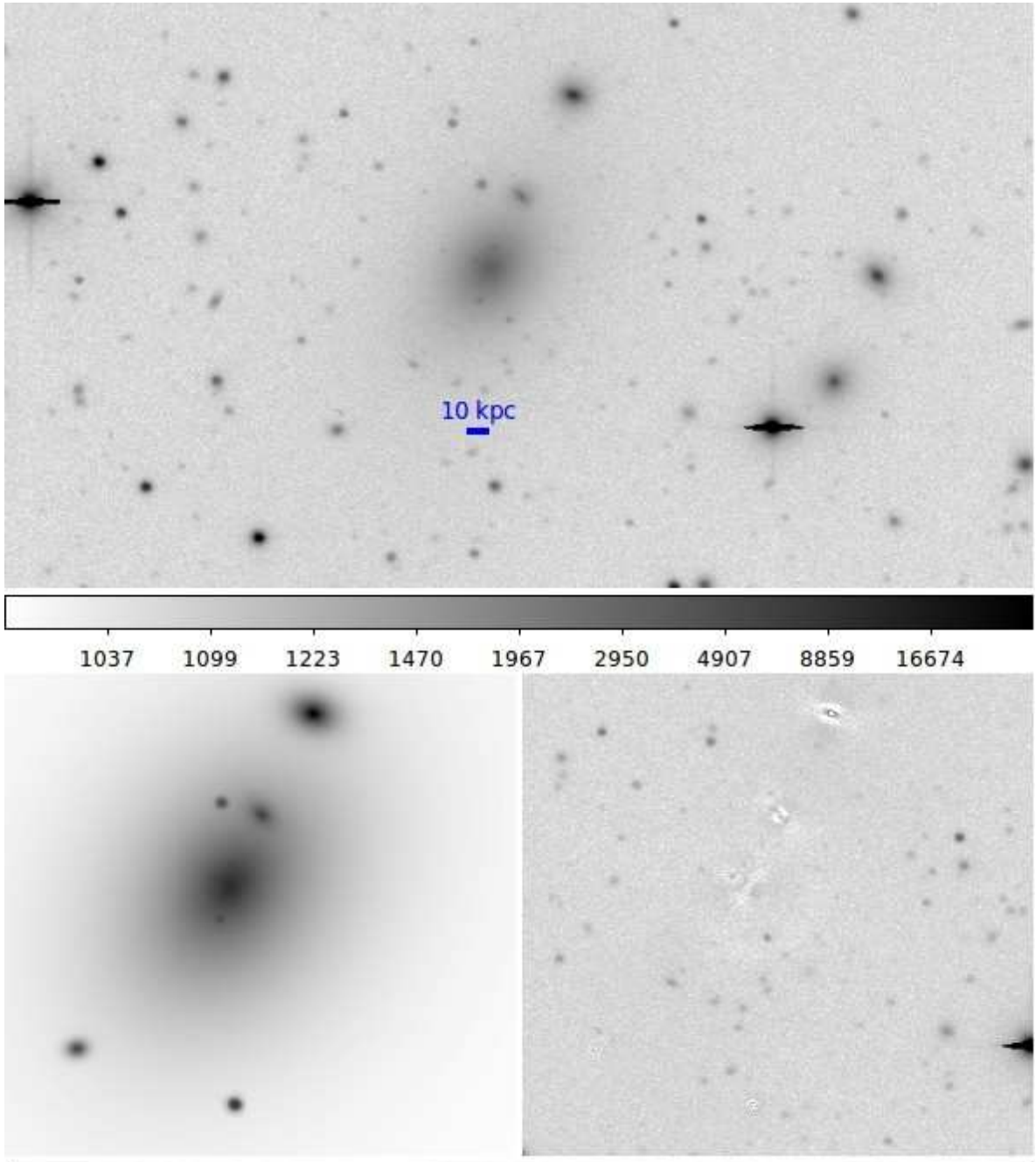}
\caption{R-band LOCOS image of Holm~15A. 
  The upper panel shows a section of the original
  image with a 10~kpc scale bar ($1\arcsec=1.074\,\mathrm{kpc})$. The
  lower left panel shows the GALFIT Nuker-law model of Holm~15A 
  with the single-S\'ersic models for eight neighboring galaxies. The
  lower right panel depicts the residual image. The central horizontal
  bar shows the intensity scale in arbitrary units. 
  \label{fig1}} 
\end{figure}

\subsubsection{Photometry} 
The Nuker law, introduced by  \citet{La95}, is a ``broken" power law  given by
\begin{equation}\label{Eq1}
 I(r)= 2^{(\frac{\beta-\gamma}{\alpha})} I_{b}
       \left(\frac{r_{b}}{r}\right)^{\gamma} 
       \left[ 1 + \left(\frac{r}{r_{b}}\right)^{\alpha}\right]^{(\frac{\gamma - \beta}{\alpha})}
\end{equation}
where $\gamma$ is the power index of the inner cusp,  $\beta$ is that
of the envelope, $\alpha$ is that at the ``break radius"  $r_{b}$, the
radius of maximum curvature in $(\log I, \log r)$ space, and
$I(r_{b})=I_{b}$ is the intensity at the break radius. 
$r_b$ has been used as a scale indicator, but the ``cusp radius''
$r_{\gamma} \equiv r_{b} \left(\frac{\frac{1}{2} - \gamma}{\beta -
  \frac{1}{2} } \right)^{\frac{1}{\alpha}}$, 
the radius where $\scriptstyle{ \left(\case{d\log I}{d\log r} \right)=-\left(\onehalf\right)}$,
correlates better with other galaxy properties
\citep[e.g.,][]{Ca97,La07a}.

GALFIT was used to fit Holm~015A with an elliptical generalization of
Eq.~\ref{Eq1} including centroid, axis ratio, and position angle
(i.e., nine free parameters in total), simultaneously with
single-S\'ersic fits to eight neighboring galaxies. The input
parameters for those eight galaxies  were taken from a 
DGCG run over the whole  LOCOS image.  
The star  SDSS J004159.70-091937.3 ($\mathrm{m_{R}}=15.85$ mag)
$2\farcm7$ from the center of Holm~15A was selected to generate the
local PSF for Holm~15A. GALFIT proceeds by convolving a local PSF with
the  model component on a pixel center and minimizes $\chi^2$
against the galaxy of interest data 
\citep[see][for details]{RH01,Peng10}.  The sky background 
was modeled without including gradients.  In simulated data it has
been found that fitting neighboring  bright objects simultaneously
with the galaxy of interest gives more accurate results than simple
pixel masking  \citep[e.g.,][A\~norve \& L\'opez-Cruz 2014]{Ha07}.
Since the background is the most important source of uncertainty in
surface brightness modeling, 
we carefully investigated how our model parameters depend
on the background level, using the combination of  one to five
S\'ersic models to remove the light of Holm~15A with other objects
masked. For each (fixed) value of the background, a new fit to (1) was 
generated. 
This allowed us to make a realistic assessment of the errors
for each fitted parameter \citep[cf.,][]{HH13}. The results are
given in Table~\ref{tbl-1}. The large core size of Holm~15A,
$r_{\gamma}=4\farcs26 \pm 0\farcs06$, and its low redshift allow
accurate parameter estimates even under the modest seeing conditions
of the LOCOS frame: the value of $r_b$ is more than ten times larger
than the seeing FWHM. The original LOCOS image, the model, and the
residual are shown in Fig.~\ref{fig1}. 
 
We confirmed our result using the MENeaCS image. The fitting strategy 
was the same as for the LOCOS, except that we found a slight
background gradient, and did not perform such a detailed error
analysis. Despite differences in telescope size, CCD
format, pixel resolution, seeing, waveband, etc., the
resulting parameter set (Table~\ref{tbl-1}, second row) agrees well with the LOCOS 
result. The agreement of the fitted values of
$r_\gamma$ and $\gamma$ is noticeable. The inner slope of Holm~15A is flat ($\gamma=0$,  see Table.~\ref{tbl-1}) but  in less luminous galaxies $\gamma$ tends to be larger  \citep[e.g.,][who applied GALFIT to large sample of early type galaxies (ETG)]{RH01}.  We note that the value of the outer slope $\beta$, found from
either fit is the largest ever reported --- compare the mean and
maximum of $\bar{\beta}=1.4\pm0.2$ and $\max(\beta)=2.63$ in
\citet{La03}.
 
\begin{figure}
\begin{center}
\includegraphics[clip,angle=90,scale=0.80]{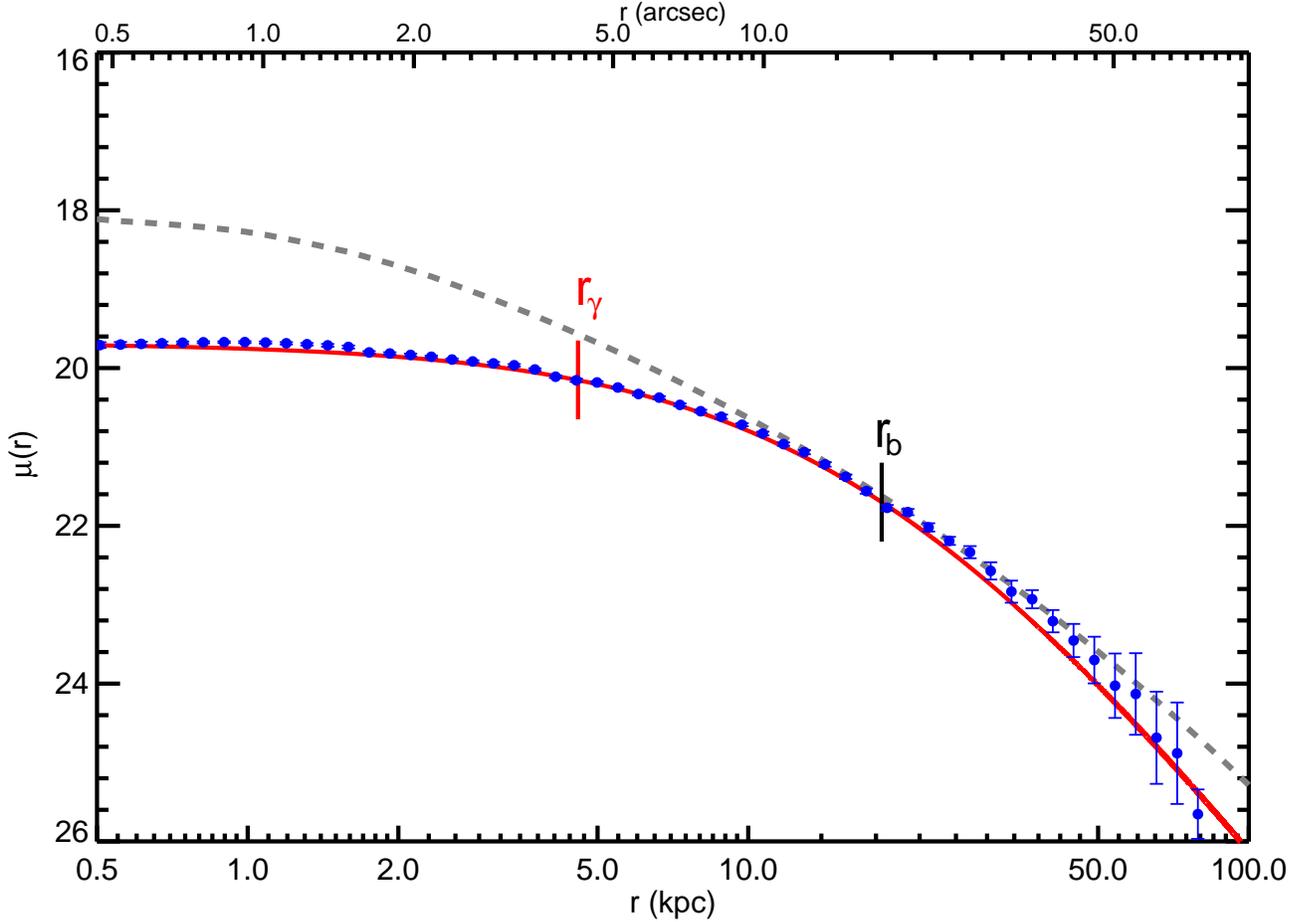}
\caption{ 
Two independent computations of the surface-brightness profile of
 the LOCOS image of  Holm~15A in the $R$-band. The blue dots, marked
 with error bars, were generated using IRAF {\em ellipse}  on a
 seeing-deconvolved image of Holm~15A while the red continuous line
 is the profile generated after a GALFIT Nuker fit to a direct image
 of Holm~15A (see Table~\ref{tbl-1}, and \S2.1.1 for details). The
 agreement between 1D and 2D surface brightness modeling is
 excellent.  A de Vaucouleurs profile (grey dashed
 line) represents well the light distribution in the outskirts, but
 over-predicts the surface brightness inside the break radius
 $r_{b}=18.48 \pm 0.04\;\mathrm{kpc}\: (17\farcs21 \pm 0\farcs04)$
 (black bar). The cusp radius $r_\gamma=4.57 \pm 0.06\; \mathrm{kpc}$
  is also indicted (red bar).  \label{fig2}}  
 \end{center}

\end{figure}

 We used the IRAF {\em ellipse} package on the LOCOS image to derive a
1D surface brightness profile of Holm~15A (blue points with errors in Figure~\ref{fig2}).
We worked on a deconvolved image (the PSF was generated using neighboring stars)  generated using STSDAS Maximum Entropy Method (MEM) inversion, with errors calculated by applying {\em ellipse} to 50 simulated images, where the background was randomly varied $\pm 1\, \sigma$,  pixel-by-pixel.  We compared deconvolved images using  MEM and the Richarson-Lucy approach (e.g., PL12), we found  negligible differences.    From this 1D profile we can determine $r_b$ and $r_{\gamma}$ directly using their definitions (given above). These non-parametric (${}^{np}$) estimates are $r_{b}^{np}=17\farcs4\pm0.2\; (18.7\pm0.2\, \mathrm{kpc)}$ and  $r_{\gamma}^{np}=4\farcs27\pm 0\farcs05\;(4.59\pm0.05\,{\rm kpc})$, these agree to better than one pixel with the results in Table~\ref{tbl-1}. To compare the results from 1D and 2D surface brightness modeling,  we use the parameters resulting after the GALFIT Nuker Law fit to  Holm~15A, whose parameters are given in Table \ref{tbl-1} (first row). The resulting profile (red continuous line in Fig. \ref{fig2}) closely matches the blue dots, indicating excellent  agreement between 1D and 2D modeling.   We conclude that the core scale is reliably measured by either a parametric or non-parametric approach.

The total stellar light in Holm~15A was modeled by the sum of two
S\'ersic profiles.  We obtained a total absolute magnitude
$M_{V}^{\mathrm{total}}=-23.81\pm0.10\; (m_V^{\mathrm total}=13.28)$
\citep[cf.,][]{DMM11}, where we have assumed $(V-R)=0.61$, as is
typical for giant early ETG. We attempted to fit a de Vaucouleurs profile
(dVP) to the LOCOS image (using GALFIT) and the 1D profile. 
This function rises too steeply at small radii and gives large
residuals inside $r_b$ (Fig.~\ref{fig2}). We estimate the {\em missing
light} in this fit by comparing the dVP integrated magnitude
$M_{V}^{\mathrm{dVP}}=-24.9^{+0.3}_{-0.4}$, where the errors are
estimated by comparing  the 1D and 2D dVP fits  to Holm~15A, 
with the value of $M_{V}^{\mathrm{total}}$ \citep[an approach similar
  to that advocated by][]{KH13}.  Hence, the approximate luminosity
missing at $r < r_b$ relative to the dVP is $L_{V,def} \simeq 5^{+3}_{-2} \times 10^{11}\,L_{V\sun}\,(M_{V\sun}=4.83)$. Missing light 
has been found to correlate with SMBH mass \citep[e.g.,][]{KB09, KH13}

\subsubsection{Spectroscopy}

By applying Fourier Quotient and Cross-Correlation methods, and the
synthesis code STARLIGHT \citep{CF05}, on the SDSS DR9 spectrum of
Holm~15A we obtain velocity dispersions of $226 \pm 50\; \mathrm{km\,s^{-1}}$  and 
$305 \pm 15\;\mathrm{km\,s^{-1}}$. While SDSS DR9 reports a velocity
dispersion of $322 \pm 13\; \mathrm{km\,s^{-1}}$, \citet{FIF95}
reported $289 \pm 31\; \mathrm{km\,s^{-1}}$ and a slight velocity
gradient. Based on these values, our best estimate for the velocity
dispersion is ${\sigma_{gal} = 310 \pm 15 \ \rm km \, s^{-1}}$, or
about $80 \ \rm km \, s^{-1}$ less than 
the velocity dispersion of  A2261-BCG ($\sigma=387\pm16 \; \mathrm{km\,s^{-1}}$, PL12). 

SDSS DR9 reports coordinates $\alpha_{2000}=00^{\mathrm h} \,41^{\mathrm m}\,50\fs46$
$\delta_{2000}=-09\arcdeg \,18\arcmin \,11\farcs34$ for the 3-arcsec diameter
fiber on Holm~15A. We used this to search for lines associated with
AGN activity, by correcting for extinction, removing stellar continuum
using STARLIGHT \citep[as in][]{Pa12}, and fitting Gaussian line
profiles. We found line ratios
\begin{displaymath}
\scriptstyle{
\begin{array}{ll}
\log\left(\case{[{\rm O \,III}]\,\lambda5007}{{\rm H}_{\beta}}\right) = 0.013 & \log\left(\case{[{\rm N\,II}]\,\lambda6584}{{\rm H}_{\alpha}}\right) = 0.302  \\
\log\left(\case{[{\rm O\,I}]\lambda6300}{{\rm H}_{\alpha}}\right) = -0.567 & \log\left(\case{[{\rm S\,II}]\lambda\lambda6717,31}{{\rm H}_{\alpha}}\right)= -0.014
\end{array}
}
\end{displaymath}
with mean errors of about $\pm0.08$ \citep{CP11}.

\subsection{Infrared Data}
We used a 2MASS $K_s$ image to determine the integrated infrared
luminosity of Holm~15A. Using elliptical apertures we determined a
total (asymptotic) magnitude  
$m_{K_s}=10.08\pm0.03$ \citep[cf.,][]{2MASS06}, so that
$M_{K_s}=-26.76\pm0.03$~mag and 
$L_{K_s}=(1.03 \pm 0.03)\times10^{12}\;L_{K_{s}\sun}\,
(M_{K_{s}\sun}=3.27)$. We find $(V-K_{s})_0 = 3.2$ for Holm~015A, this
color falls within the expected range for luminous ETG
\citep[e.g.,][supplemental material]{KH13}, and hence no correction is
needed to the standard 2MASS photometry.  

\subsection{Radio and X-ray Observations}

The radio field of Abell~85 is complicated, but our analysis of archival
VLA datasets clearly distinguishes radio emission from Holm~15A from
other radio components associated with \objectname{0039-095B}
\citep{Owen84}, and the radio structure associated with the cluster
merger \citep{slee}. Using the AR286 dataset (from 1992~December), which provides the best angular resolution available, we find a compact (0.39~arcsec major axis) radio source with peak flux density
$S_{8.4\,\mathrm {GHz}} = 0.95 \pm 0.05 \ \mathrm{mJy}$ and integrated
flux density $S_{8.4\,{\rm GHz}} = 1.8 \pm 0.1 \ \rm mJy$ at
$\alpha_{2000}= 00^{\mathrm h} \, 41^{\mathrm m}  \,  50\fs   471\,\pm0\fs001, \,                                                                                                  
  \delta_{2000}=-09\arcdeg \, 18\arcmin \,11\farcs42 \,\pm0\farcs01$
associated with the core of Holm~15A. This component of an overall 6.5-mJy source, about 5~arcsec in size, appears to be 
extended along a direction parallel to the major axis of the
galaxy. The 8.4-GHz structure at the centre of the galaxy lies within a diffuse structure $\approx 
15$~arcsec in size at 1.4~GHz, with an NVSS
flux density $S_{1.4\,{\rm GHz}} = 56.7 \pm 2.5 \,{\rm mJy}$
\citep{NVSS98}. Adopting a spectral index $\alpha=1.08$
\citep{Bu90}, we obtain a radio power  $P_{1.4\, {\rm GHz}}= (4.2                                                 
\pm 0.2) \times 10^{23}\, {\rm W\,Hz^{-1}}$ for 0039-095B.

We analyzed {\em Chandra} archival data (OBSID 904) using CIAO
analysis task {\em wavdetect},  and   found no X-ray point source near
the center of  Holm~15A.  However, there is a compact  X-ray source,
which appears extended although with a bright center, associated with
the position of \objectname{SDSS~J004150.75-091824.3} (hereafter
J004150). This source, $13\farcs74$ ($< r_b$), from the center of
Holm~15A, is a quasar candidate with $z_{phot}\sim0.9$ \citep{RQ09}. 

\section{Discussion}
\begin{figure}
\epsscale{.90}
\plotone{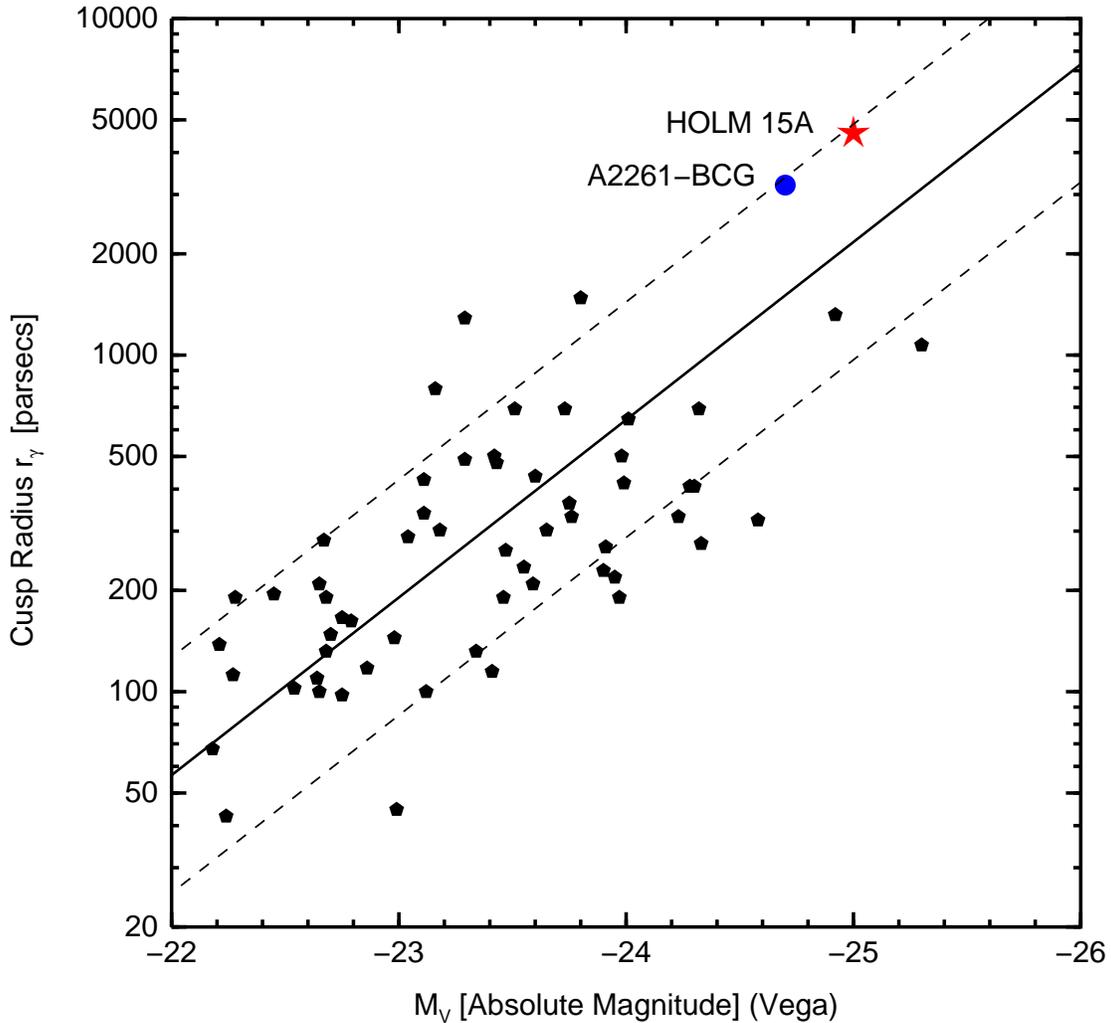}
\caption{The correlation between $r_{\gamma}$ and luminosity in the
  $V$-band for core BCGs \citep[data taken from][black
  pentagons]{La07a}, with Holm~15A and A2261-BCG marked by
  the red star and blue  dot, respectively. 
  Holm~15A is about 1~kpc larger than A2261-BCG. The
  solid line is the fit given by  Equation~4 of \citet{PL12}. Dashed lines represent the scatter
  about the mean of the correlation ($1\sigma$ = 0.35 dex). Both
  A2261-BCG and Holm~15A fall within $1\sigma$ on the high side of
  the correlation.
\label{fig3}}
\end{figure}

We have used optical images taken with different
telescope configurations, using parametric and non-parametric approaches,  and
consistently find a large cusp radius for Holm~15A ($r_{\gamma} = 4.57\pm0.06 \ \rm kpc$).   This value of the cusp radius makes Holm~15  about $1\, \mathrm{kpc}$ larger than A2261-BCG
($r_{\gamma}=3.2\, \pm 0.1$ kpc), hitherto the largest-cored BCG (PL12). 
Holm~15A's cusp radius is about $18 \times$ the mean cusp
radius for cored BCG \citep[based on data from][]{La03}. 
Figure \ref{fig3} shows that although Holm~15A is the largest cusp-BCG, it nevertheless lies less than $1\sigma$ above the correlation between $r_\gamma$ and luminosity (as does A2261-BCG). This might
suggest that a common mechanism is responsible for the formation of cores over a wide range of scales.

The line ratios (\S2.1.2) suggest that Holm~15A is a LINER
\citep[e.g.,][]{Ho08,Pa12}. By comparison, A2261-BCG is devoid of
emission lines (PL12). Low emission from 
molecular gas \citep{SC03} and dust \citep{Q08} 
suggests that there is little star formation in Holm~15A.  
Indeed, the radio power of Holm~15A (\S2.2), being larger than $P_{1.4\, {\rm GHz}}=
10^{22.75}\, {\rm W\,Hz^{-1}}$, cannot be explained by star formation
alone \citep{Mo03}. Hence, we conclude
 that optical line emission and radio continuum are dominated by AGN activity.  
 A2261-BCG is  a radio-AGN, it  is slightly
more luminous than Holm~15A ($P_{1.4\, {\rm
 GHz}}= 5\times 10^{23}\, {\rm W\,Hz^{-1}}$, PL12)

Using known scaling relations we formed  the five SMBH
mass estimates  given in Table~\ref{tbl-3}.  The
mass inferred from the $\mathrm{M}_{\bullet}-\sigma$ relation is 
the lowest. This might be a consequence of the
breakdown of the $\mathrm{M}_{\bullet}-\sigma$ scaling law for
$\sigma \gtrsim 270\; \mathrm{km\,s^{-1}}$
\citep[e.g.,][]{La07a,KH13}. Three scaling relations
suggest BH masses $\mathrm{M}_{\bullet} \gtrsim 10^{11}
\;\mathrm{M_{\sun}}$, but these relations have large scatter and were
derived with limited samples. Therefore, we conservatively
suggest that Holm~15A hosts an SMBH with $\mathrm{M}_{\bullet} \sim
10^{10}\;\mathrm{M_{\sun}}$.  


An SMBH has a strong stellar-dynamical influence within radius
$r_{f}=\left(\case{G\mathrm{M_{\bullet}}}{\sigma^2}\right)$. If we adopt
$\mathrm{M}_{\bullet} = 10^{10}\; \mathrm{M_\sun}$ and
velocity dispersion 
$\sigma \approx 310 \,\mathrm{km\,s^{-1}}$, 
$r_{f} \thicksim 450\,\mathrm{pc}\;(0\farcs42)$. Thus only with the
largest mass estimate from Table~\ref{tbl-3} can we
interpret the cusp radius as due to the gravitational influence of an SMBH
causing $r_f\thicksim r_{\gamma}$.  

Let's suppose that Holm~15A contains an SMBH binary with
total mass $\mathrm{M_{\bullet_B}}$, then we would expect the separation between the 
SMBH components to be
$a_{B} \thicksim \left(\case{G \mathrm{M_{\bullet_B}}}{2\sigma^2}\right)
         \thicksim \left(\case{r_f}{2}\right)$
\citep{DSD12}. For an SMBH binary  total mass $\mathrm{M_{\bullet_B}}=10^{10}\;
\mathrm{M_\sun}$ this implies $a_{B}\thicksim 225
\,\mathrm{pc}\;(0\farcs21)$, which is close to the average cusp
scale of BCGs, but much less than the scale of Holm~15A.

The morphology of the radio emission in Holm~15A is not clear in
current data, since most of the flux density is in rather diffuse
structure, which could be much distorted. If this is the case, then by
analogy with RBS~797 \citep{GG13}, a high-sensitivity,
high-resolution, map of the center of Holm~15A could
test whether Holm~15A hosts an SMBH binary.

J004150 appears to be at far higher redshift than Holm~15A. However,
if the light of Holm~15A was not properly modeled, the photometric
redshift estimated for J004150 could be in error. Moreover, if J004150 is
a quasar at $z = 0.9$, then extended X-ray emission from the hot interstellar
medium is unlikely to have been detected in the current {\em Chandra} exposure. 
Instead, the apparent X-ray extension is more consistent with a low-$z$ AGN. 
It is possible that J004150 is a third SMBH component associated with Holm~15A. This
can be tested by optical spectroscopy of J004150.

We follow \citet[][\S 6.10]{KH13} to estimate the mass of the SMBH associated to the cluster dark matter halo.   
Abell~85 has velocity dispersion $\sigma_{cl}=752\pm34\;\mathrm{km\,s^{-1}}$ (from our own analysis and
the literature) and hence should host a central
$\mathrm{M}_{\bullet} \thicksim 1.5\times10^{11}\;\mathrm{M_{\sun}}$
SMBH. Holm~15A is the central galaxy of Abell~85, and three of the 
mass estimates in Table~\ref{tbl-3} are consistent with the expected
cluster central SMBH mass. Such agreement is unexpected unless dark
matter halos are scale-free, and the SMBH-dark matter coevolution 
is independent from  the effects of  baryons \citep[cf.,][]{KH13}. Otherwise, we should accept that
ultramassive ($\mathrm{M}_{\bullet} \geq 10^{10}\;\mathrm{M_{\sun}}$) BH in BCG
 follow special scalings  as suggested by \citet{Hla12}, and  that  the masses of   SMBH cannot grow indefinitely \citep{NT09}.

\section{Conclusions}
We have found that Holm~15A has the largest core known so far, with
$r_{\gamma} = 4.57\pm0.06 \, \mathrm{kpc}$. A central AGN supports the
presence of a central BH, which could be ultramassive. Very large
cores ($r_{\gamma} \gtrsim 1\, \mathrm{kpc}$) seem to be rare (see
Figure~\ref{fig3}) and may represent a relatively brief phase in
the evolution of BCG, as calculated merging times for SMBH binaries appear to be 
relatively short \citep{Ka14}. Other mechanisms might be at work, because  time scales for cusp regeneration seem too long \citep[e.g.,][]{Me06}. If SMBH growth is regulated by galaxy mergers \citep[e.g.,][]{Me06, BS11}, their
final masses were set, perhaps, by initial conditions
\citep[e.g.,][]{Tre13}.    


The physical condition of Holm~15A may represent one of the best laboratories for testing the SMBH ``scouring''
scenario for the creation of BCG cores. Follow-up observations
that might test for the presence of a second, or third, SMBH component
include improved radio mapping, optical spectroscopy of J004150, and stellar-
and gas-dynamical mapping using HST or a large ground-based telescope
\citep[e.g.,][]{DB09, Mc12} to investigate velocities on scale
$0\farcs4$. Since only a few BCGs have dynamically-determined
BH masses \citep[e.g., Table 2 in][]{KH13},
further studies on Holm~15A (and A2261-BCG) could 
provide  crucial  tests  of  the applicability of SMBH mass scaling laws, the core ``scouring" scenario, and, hence, on the coevolution of BH.

\acknowledgments


Funding for SDSS-III has been provided by the Alfred P. Sloan Foundation, the Participating
Institutions, the NSF, and the U.S. Department of Energy Office of Science. We thank  Chien Peng for comments and advice.

\clearpage

\begin{deluxetable}{cccccccccc}
\tabletypesize{\scriptsize}
\tablecaption{Holm~15A: Nuker law fits   \label{tbl-1}}
\tablewidth{0pt}
\tablehead{
\colhead{$\mu_{b}\,\mathrm{\left[\frac{mag.}{arcsec^2}\right]}$} &
\colhead{$r_{b}\,\mathrm{[arcsec]}$}& 
\colhead{$r_{b}\,\mathrm{[kpc]}$} &
\colhead{$\alpha$} & 
\colhead{$\beta$}& 
\colhead{$\gamma$}&
\colhead{$r_{\gamma} \mathrm{[kpc]}$}&
\colhead{$e$} &
\colhead{P. A.} & 
\colhead{Data} 
}
\startdata
$21.78\pm 0.01$&$17.21\pm 0.04$ & $18.48\pm 0.04$ &$1.24\pm 0.01$ & $3.33\pm 0.02$ & $0.0\pm0.0$ &$\mathbf{4.57\pm 0.06}$ & 0.26 &-33.33 & \tablenotemark{\bigstar}\\
22.32&19.09 & 20.50   & 1.22& 3.62 & 0.0& $\mathbf{4.57}$& 0.24&-34.07&\tablenotemark{\bullet}\\
\enddata
\tablenotetext{~}{COLUMNS-- 1: surface brightness,  2: break radius, 3: break radius, 4: power index at $r_b$, 5: outer  power index, 6: inner power index, 7: cusp radius in kpc, 8: ellipticity, 9: position angle in degrees,10: Data Source.}
\tablenotetext{\bigstar}{~LOCOS \citep{L00}, Telescope: KPNO 0.9m, CCD: T2KA; pixel scale: $0\farcs68$/pixel. filter: $R$ (Kron-Cousins), exposure time: $900\,\mathrm{s}$, seeing: $1\farcs6$ FWHM, FOV: $23\farcm 2 \times 23\farcm2$. }
\tablenotetext{\bullet}{~MENeaCS \citep[][]{Sa11}, Telescope: CFHT 3.5m, CCD: MegaCam; pixel scale: $0\farcs187$/pixel, filter: $r^{\prime}$ (SDSS), exposure time: $120\, \mathrm{s}$, seeing: $0\farcs74$ FWHM, FOV: $0\fdg96 \times 0\fdg94$.}
\tablecomments{centroid: $\alpha_{2000} = 00^{\mathrm h}  \, 41^{\mathrm m} \, 50\fs467$,  $\delta_{2000}=-09\arcdeg \, 18\arcmin \,11\farcs57$ }

\end{deluxetable}



\begin{deluxetable}{lcl}
\tablecaption{Holm~15A: Black Hole Mass Estimates\label{tbl-3}}
\tablewidth{0pt}
\tablehead{
\colhead{Relation} & \colhead{$\mathrm{M}_{\bullet}\; [\mathrm{M}_\sun]$}  &\colhead{Reference} 
}
\startdata
$\mathrm{M}_{\bullet}-\sigma$&$\thicksim2.1\times 10^{9}$&
\citet[][Eqs. 6]{KH13}\\
$\mathrm{M}_{\bullet}- L_{K, bulge}$\tablenotemark{*} &$\thicksim9.2\times 10^{9}$&\citet[][Eqs. 7]{KH13}\\
$\mathrm{M}_{\bullet}-L_{V, def}$&$\thicksim2.6\times 10^{11}$ & \citet[][Eq. 3]{KB09}\\ 
 $\mathrm{M}_{\bullet}-r_b$&$\thicksim1.7\times 10^{11}$&
  \citet[][Eq. 13]{Ru13}\\
   $\mathrm{M}_{\bullet}-r_{\gamma}$&$\thicksim3.1\times 10^{11}$&\citet[][Eq. 26 ]{La07a} \\
 \enddata
 \tablenotetext{*}{Taking the entire galaxy
as a classical bulge, and correcting the value of $\mathrm{H}_0$}
\end{deluxetable}

\end{document}